\documentclass[10pt]{iopart}
\usepackage{svg}
\usepackage{bm}
\usepackage{citesort}

\usepackage{graphicx}
\usepackage{dcolumn}
\begin{document}

\title[Neural networks and kernel ridge regression for excited states dynamics of CH$_2$NH$_2^+$]{Neural networks and kernel ridge regression for excited states dynamics of CH$_2$NH$_2^+$: From single-state to multi-state representations and  multi-property machine learning models}


\author{Julia Westermayr$^1$, Felix A. Faber$^2$, Anders S. Christensen$^2$, O. Anatole von Lilienfeld$^2$, and Philipp Marquetand$^1$}

\address{$^1$ University of Vienna, Faculty of Chemistry, Institute of Theoretical Chemistry, W\"ahringer Str. 17, 1090 Wien, Austria}
\address{$^2$ Institute of Physical Chemistry and National Center for Computational Design and Discovery of Novel Materials (MARVEL), Department of Chemistry, University of Basel, Klingelbergstr. 80, CH-4056 Basel, Switzerland}

\ead{anatole.vonlilienfeld@unibas.ch}
 \ead{philipp.marquetand@univie.ac.at}
\vspace{10pt}
\begin{indented}
\item[]August 2019
\end{indented}

\begin{abstract}
Excited-state dynamics simulations are a powerful tool to
investigate photo-induced reactions of molecules and materials and
provide complementary information to experiments. Since the applicability of these simulation techniques is limited by
the costs of the underlying electronic structure calculations, we develop and assess different machine learning models for this task. The machine learning models are trained on {\emph ab initio} calculations for excited electronic states, using the methylenimmonium cation (CH$_2$NH$_2^+$) as a model system.
For the prediction of excited-state properties, multiple outputs are desirable, which is straightforward with neural networks but less explored with kernel ridge regression. We overcome this challenge for kernel ridge regression in the case of energy predictions by encoding the electronic states explicitly in the inputs, in addition to the molecular representation. We adopt this strategy also for our neural networks for comparison. Such a state encoding enables not only kernel ridge regression with multiple outputs but leads also to more accurate machine learning models for state-specific properties. An important goal for excited-state machine learning models is their use in dynamics simulations, which needs not only state-specific information but also couplings, i.e., properties involving pairs of states. Accordingly, we investigate the performance of different models for such coupling elements. Furthermore, we explore how combining all properties in a single neural network affects the accuracy. As an ultimate test for our machine learning models, we carry out excited-state dynamics simulations based on the predicted energies, forces and couplings and, thus, show the scopes and possibilities of machine learning for the treatment of electronically excited states.
\end{abstract}
\maketitle
\ioptwocol
\section{\label{sec:level1}Introduction}
Many fundamental processes in nature and life are direct consequences of excitation of molecules by light. 
For example, photosynthesis~\cite{Vass2007ANAS}, vision with photo-receptors in the  eye~\cite{Schoenlein1991S,Garavelli1999JACS}, or the root cause of diseases such as skin cancer~\cite{Schreier2007S,Rauer2016JACS} are all based on a photo-induced process.
Understanding the dynamics and kinetics of compounds undergoing excitation can therefore provide insight into why and how these processes occur, or can be used to help designing new drugs~\cite{Ahmad2016IJP} or materials~\cite{Haese2016CS,Butler2018N,Liu2015JPCL}.

As diverse and important photo-induced processes are, as challenging they are to study.
After a molecule is irradiated with light, it enters a higher electronic state, from which several processes can occur.
In general, one can distinguish between radiationless and radiative transitions from one electronic state to another.
Concerning the former, internal conversion between states of same spin multiplicity or intersystem crossing between singlet and triplet states, can lead to a fast redistribution of energy. In this way, the photostability of many molecules can be explained. However, if the energy redistribution is not complete, a substantial amount of energy is stored in the molecule and can provide the starting point of chemical reactions, e.g.~giving rise to photodamage~\cite{Domcke2013NC,Liu2017SR,Mai2018WCMS}.
From a theoretical point of view, corresponding reaction channels and kinetic models can be obtained via excited-state molecular dynamics simulations, that can be directly compared to experiments.
Moreover, a detailed insight into the electronic and structural properties that determine the dynamics of a molecule can be gained.

In order to study photodynamics, knowledge about the high-dimensional electronic potential energy surfaces (PESs) of a molecule, on which the nuclei are thought to move classically or quantum mechanically, has to be provided.
Since the computation of PESs in advance is a rather tedious task, on-the-fly {\emph ab initio} molecular dynamics is usually the method of choice.
In terms of accuracy and computational efficiency, mixed-quantum classical methods -- such as the surface-hopping methodology that is used in this work -- are often a good compromise.
Despite overcoming the Born-Oppenheimer approximation, the electronic and nuclear motions are treated differently in surface hopping, the electrons with quantum mechanics and the nuclei with classical mechanics. In this way, large molecules, i.e., with up to hundreds of atoms, can still be treated.
Still, a lot of electronic structure calculations are necessary and the latter represent an important bottleneck limiting the simulation times of nonadiabatic dynamics to the range of femto- to picoseconds~\cite{Mai2016JPCL,Doltsinis2006NIC,Subotnik2016ARPC,Marquetand2017M}.

Recently, with the rise of machine learning (ML), efficient yet accurate models have been developed to tackle this problem and partially replace quantum chemical calculations in molecular dynamics simulations. The main advantage of ML models is that, at least in principle, they can predict any molecular property, typically with much improved efficiency when compared to their quantum chemical counterpart. This can be achieved by learning the relation between a molecular structure (represented by some translation- and rotation-invariant representation) and some target property (provided by quantum chemistry, usually real-valued or complex numbers). A trained ML models can infer quantum chemistry results without suffering significant accuracy loss, assuming an interpolating regime and that enough data points are provided for training~\cite{Behler2015IJQC,Rupp2015IJQC,Gastegger2016JCP,Hansen2013JCTC}. Successful application of ML models for dynamics simulations in the electronic ground state~\cite{Shen2018JCTC,Wang2017PCCP,Gastegger2017CS,Li2015PRL,Botu2017JPCC,Artrith2017PRB,Behler2017ACIE,Chmiela2017SA} or excited states~\cite{Behler2008PRB,Carbogno2010PRB,Dral2018JPCL,Hu2018JPCL,Xie2018JCP,Chen2018JPCL,Guan2019PCCP,Westermayr2019CS} already exist, and show their potential for speeding up molecular dynamics simulations.

However, excited-state {\emph ab initio} molecular dynamics simulations still face another problem due to the fact that not only one potential energy surface has to be treated, but several electronic states including the couplings between them, that should be included in the ML model as well~\cite{Westermayr2019CS}.
These are nonadiabatic couplings (NACs) between states of same spin multiplicity and spin-orbit couplings between states of different spin multiplicity, e.g., singlet and triplet states~\cite{Mai2018WCMS}. In this work, only singlet states are considered, hence only NACs will be discussed. NACs are large, when two states are in close proximity to each other and are almost zero elsewhere. At conical intersections (CIs), where two states are degenerate, NACs show singularities, which leads to a break-down of the underlying Born-Oppenheimer approximation~\cite{Doltsinis2006NIC}. These characteristics make it not only challenging to model NACs with ML, but also to converge a quantum chemistry calculation near CIs. However, CIs are of huge importance for dynamics simulations since the probability for transitions between different states rises the closer a molecule moves towards a CI~\cite{Nanbu2010CS}. It is thus indispensable to model NACs correctly, especially in those regions. A trend can be obtained from independent works that applied kernel ridge regression (KRR) and neural networks (NNs) to replace quantum chemical calculations in nonadiabatic molecular dynamics simulations. Models based on KRR were not able to predict couplings sufficiently accurate to simulate the nonadiabatic molecular dynamics of 6-aminopyrimidine~\cite{Hu2018JPCL} and an adiabatic spin-boson Hamiltonian model~\cite{Dral2018JPCL}. Therefore, quantum chemical computations in regions close to CIs had to be performed, and could not be substituted by ML predictions. 
In contrast, artificial neural networks (NNs) were able to reproduce couplings also in such critical regions of the PES and provided a model to completely replace quantum chemistry for excited state molecular dynamics of different molecules~\cite{Xie2018JCP,Chen2018JPCL,Guan2019PCCP,Westermayr2019CS}.

This observation raises the question why some models fail for some excited-state properties and why others do not. Yet no studies are available that provide a detailed insight into the underlying problems of excited-state properties that need to be tackled with ML models. We therefore seek to investigate relevant properties for nonadiabatic molecular dynamics simulations by using KRR and NNs. We consider energies, corresponding gradients, as well as NACs for the methylenimmonium cation, CH$_2$NH$_2^+$. We further investigate (transition) dipole moments, since they are important, for instance, for the assessment of biological properties or infrared spectra intensities~\cite{Harris1989} and are often targeted with ML models lately~\cite{Ramakrishnan2014SD,Huang2016JCP,Gastegger2017CS,Schuett2018JCP,Nebgen2018JCTC,Sifain2018JPCL,pereira2018JC,Schuett2019,Schuett2019JCTC,Christensen2019JCP}. We address the questions, which ML model gives the best performance for which property, which molecular representation is best suitable for which algorithm and property and how can different properties be predicted more accurately. Answers are obtained via learning curves and scatter plots between reference data and predicted data. Especially learning curves are powerful tools to assess the learning efficiency of a ML model by plotting the prediction error against the training set size, $N$, in logarithmic scale~\cite{Corinna1994,Mueller1996NC,vonLilienfeld2018ACIE,Christensen2019JCP}.
By finding the best solution for each problem, the goal is to find out how and with which ML model to best treat different excited state properties and to get a better insight into their underlying pattern. 
The most promising models are used to carry out surface-hopping nonadiabatic molecular dynamics simulations. 

\section{Methods}
\subsection{Training set}
The training set that is used is taken from Ref.~\cite{Westermayr2019CS} and represents a conformational subspace of the methylenimmonium cation, CH$_2$NH$_2^+$.
This training set involves three singlet states and was already shown to cover the relevant conformational space that is visited during excited-state molecular dynamics simulations after excitation to the bright second excited state ($S_2$, $\pi\pi^{\ast}$). It is thus considered to constitute an optimal set for analysis of different ML models as well as common molecular representations.  
The quantum chemical reference method is the multi-reference configuration interaction method accounting for single and double excitations with the basis set aug-cc-pVDZ (MR-CISD(6,4)/aug-cc-pVDZ). The active space consists of 4 active electrons in 6 active orbitals. The data set contains 4770 data points, with results being obtained from models trained on 4000 data points. Each data point contains the xyz-coordinates of a molecular structure as well as energies for three singlet states, corresponding gradients, (transition) dipole moments, and NACs between each state. In total, 3 energy values, 54 gradient values, 27 values for (transition) dipole moments, and 162 values for NACs had to be predicted. The relation, $NAC_{ij}=-NAC_{ji}$, can be used as well as $NAC_{ii}=0$ to reduce the amount of predicted values to 54.
For more details on the training set and its generation, see Ref.~\cite{Westermayr2019CS}.

For the computation of nonadiabatic molecular dynamics using the surface-hopping method, the nuclei are treated with classical mechanics while the electrons are described by quantum mechanics. The electronic wavefunction is parametrically dependent on the nuclear coordinates $\mathbf{R}$ and is an eigenfunction of the electronic Hamiltonian, $\hat{H}_{el}(\mathbf{R,r})$. By solving the resulting time-independent electronic Schr\"{o}dinger equation,
\begin{equation}
    \hat{H}_{el}\Psi(\mathbf{r;R}) = E_{el}(\mathbf{R}) \Psi(\mathbf{r;R)},
\end{equation}
the energies of the different electronic states are obtained as the eigenvalues. For each nuclear configuration of a molecule, the electronic Schr\"{o}dinger equation can be solved resulting in the corresponding PESs. In this way, the potentials for the generation of a ML training set can be computed. 

For many situations, the Born-Oppenheimer approximation holds and the NACs between the different electronic states can be safely ignored, assuming the states to be decoupled. The obtained picture corresponds to the motion of the nuclei on a single PES made up by the electrons. However, within nonadiabatic molecular dynamics simulations, transitions from one state to another need to be taken into account. Those transitions, or hops, usually take place in the vicinity of a CI, where the NAC terms are not negligible anymore. The NAC vector between two states, i and j,

\begin{equation}
NAC_{ij} \approx \langle \Psi_i \mid \nabla \Psi_j \rangle,
\end{equation}

is inversely proportional to the energy gap between the two coupled states. The equation can thus be rewritten~\cite{Doltsinis2006NIC}:

\begin{equation}
\label{eq:3}
    NAC_{ij} \approx \frac{\langle \Psi_i \mid (\nabla \hat{H}_{el})\mid \Psi_j \rangle } {E_i-E_j}    \textrm{ for } i \neq j.
\end{equation}
As can be seen, if two states are degenerate, the NACs become infinitely large and similarly, they show sharp peaks around avoided state crossings. In contrast, in regions far from CIs, NAC values are almost zero. Those characteristics of NACs were already identified to be problematic for KRR~\cite{Dral2018JPCL} and NNs~\cite{Chen2018JPCL,Westermayr2019CS}. In order to get rid of the sharp spikes of NACs and to provide smoother quantities for ML models, we incorporate the energy gap of the two adjacent electronic states for learning and prediction of corresponding NACs. By using this physical relation, the accuracy of NAC predictions is supposed to increase. Another promising approach is SchNOrb, where an analytic expression for the molecular wave function is obtained, that could be powerful to evaluate NACs as the derivatives of the ML wave function~\cite{Schuett2019NC}.

In detail, for learning, the NACs obtained from quantum chemistry (labeled as QC) are multiplied with the corresponding energy gap,
\begin{equation}
C_{ij}^{\textrm{reference}} = NAC_{ij}^{QC} \cdot \mid E_i^{QC} - E_j^{QC}\mid
    \label{eq:naclearn}
\end{equation}
whereas for prediction, the learned quantity is divided by the energy gap,
\begin{equation}
NAC_{ij}^{\textrm{predict}} = \frac{C_{ij}^{ML}}{\mid E_i^{ML}-E_j^{ML}\mid}.
    \label{eq:nacpredict}
\end{equation}
The idea behind this approach is that close to a CI, where two adjacent states become degenerate, the energy gap becomes very small, hence the narrow spikes disappear and result in smoother NAC curves, which are also referred to interstate couplings~\cite{Guan2019PCCP}. In order to predict the magnitude correctly, the fitted interstate coupling values are then divided by the energy gap obtained from ML models before they are given to the surface-hopping molecular dynamics program to obtain the actual values of NACs to compute the hopping probability. 
However, this approach requires very accurate ML potentials for energies and those are also challenging to obtain in regions near CIs.

Another effect that requires special care is the random sign switch of excited-state properties along reaction coordinates that result from the wave function of two different electronic states. For example, by carrying out two computations of very similar molecular configurations, the sign of the couplings can arbitrarily switch from "+" to "-" or vice versa, making a regression model fail in such cases. This is a result of the non-uniquely defined phases of the wave functions that are arbitrarily assigned by a quantum chemistry program. In order to make the data learnable for ML models, an additional pre-processing has to be carried out, referred to as phase correction~\cite{Plasser2016JCTC,Mai2018WCMS,Akimov2018JPCL,Westermayr2019CS}.
In a similar manner as NACs, also the transition dipole moments are obtained as off-diagonal elements, according to the following expression:
\begin{equation}
    \mu_{ij} = \langle \Psi_i \mid \hat{\mu} \mid \Psi_j \rangle,
\end{equation}
and have to be phase corrected as well. The permanent dipole moments are obtained as the diagonal elements, i.e. resulting from the wave function of the same state, and can be directly used for training, since the phase enters as absolute square and cancels out.

\subsection{Surface-hopping molecular dynamics}
For surface-hopping molecular dynamics simulations with ML, the program pySHARC, a python wrapper for the SHARC (Surface-Hopping including ARbitrary Couplings) program~\cite{sharc-md2}, was used and ML models were interfaced with it. The dynamics simulations of the methylenimmonium cation were started from the same 200 initial conditions for the quantum chemical reference method and the ML models. The initial conditions were obtained from  Wigner sampling~\cite{Wigner1932PR}. A time step of 0.05 fs was selected to propagate the nuclei on the different PESs. The hopping probabilities were computed with SHARC taking the NAC vectors from ML models into account.

\subsection{Machine learning models}
In order to assess the quality of each ML model, learning curves were computed and scatter plots were analyzed.  
As ML models, NNs and KRR were chosen.
The hyperparameters of each model were optimized using 5-fold cross-validation. The data set for training and validation contained 4000 data points that were obtained by randomly shuffling the complete set of 4770 points. The rest of the data set (770 data points) was held back as a test set. For the training of the NN models, we used an early stopping mechanism to prevent overfitting. To this aim, we do not use all 4000 data points directly for training NN models, but split the training set into a training and validation set using a ratio of 9:1 as it is done in 10-fold cross-validation. The mean absolute error (MAE) of each trained model was then computed for each property on the test set, resulting in 10 similar computations. In the learning curves, we report the mean of the MAEs along with the standard deviations, which has the advantage that a measure of uncertainty can be provided. In the same way as we did for NNs, we report the MAEs of KRR models on the test set by using the mean of MAEs obtained from 10 calculations.

For the NNs, a multi-layer feed forward architecture was selected and all values of different electronic states were accounted within one NN. This means, that one NN for energies was always used to predict three energy values corresponding to the three different electronic states. 
In contrast, KRR was trained for each electronic state separately at first. In order to improve ML predictions, a representation to encode the quantum energy level was developed, allowing for the use of one KRR model for all three states at once. For KRR, the QML toolkit~\cite{QML} was used. 

We further show the influence of learning all properties with one model. This means, that we include (transition) dipole moments and NACs in addition to energies and gradients, which are already treated together in some models. This is straightforward with NNs, therefore this effect was investigated using NNs. For the gradients, we take two different approaches: Once gradients are directly trained and predicted within an individual ML model and once they are included into the training of energies and obtained as derivatives of the ML models as described in Ref.~\cite{Gastegger2015JCTC} for NNs and Ref.~\cite{Christensen2019JCP} for KRR (which is labeled as \textit{gradients derived} in the following figures).
Details on chosen parameters for KRR and NNs are given in the Supporting Information (SI) in chapter S1.1 and S1.2, respectively.

\subsection{Molecular representations}
As a molecular representation the matrix of inverse distances (inv.D.) was chosen as it was also used in Ref.~\cite{Westermayr2019CS} for NNs and in the GDML model in Ref.~\cite{Chmiela2018NC}, giving fair results. We tested the FCHL18 (Faber-Christensen-Huang-Lilienfeld) representation for KRR ~\cite{Faber2018JCP,Christensen2019JCP} as well as a development version of the FCHL19 representation~\cite{Christensen2019arXiv}, that can also be used for NNs. Strictly speaking, when we apply the FCHL19 representation and the FCHL18 representation for the treatment of gradients as response properties, kernel-based regression is carried out and no Thikonov regularization is used. For simplicity reason, we still refer to KRR in the following. If not mentioned explicitly ortherwise, FCHL18 and FCHL19 are called FCHL henceforth for KRR and NNs, respectively. 

Within the FCHL representation, each atom is described by its chemical environment by using one-, two- and three-body terms, accounting for chemical composition, distances between atoms, as well as radial contributions, respectively.  
In order to compare NNs and KRR we used the inv.D. and FCHL representations for both regressors. Gradients are treated as response properties for KKR with the FCHL representation~\cite{Christensen2019JCP} and are derived from NN potentials for energies using the inv.D. representation.

Since our ML models should predict several electronic state energies at once, we also implemented an encoding for the quantum energy level in addition to the aforementioned representations. Several possibilities were tested to describe the electronic state. For KRR, a representation for each electronic singlet state, $S=\{1,2,3\}$, containing simply numbers of 1, 2, and 3 for the three states turned out to be beneficial. Different types of representations did not result in an improved learning and only changed the additional hyperparameter, the width of the state kernel. Also for the NNs, we tested several state-encoding representations and it turned out to be best to duplicate a molecular representation $N_s$-times and multiply each copy with the corresponding state-number -- 1, 2 or 3 in this case.
 

\subsection{Kernel Ridge Regression (KRR)}

In KRR, a kernel basis function is placed on each compound (each molecule) in the training set, $\{M_k\}$, and related to a property of a query compound, $p(M)$, by:
\begin{equation}
p(M)=\sum_{k=1}^{N_m} \alpha_k K(M,M_k)
\end{equation}

with $N_m$ being the number of molecules in the training set, $K$ the kernel, and $\{\alpha_k\}$  the regression coefficients,
\begin{equation}
    \alpha=(K+\lambda I)^{-1}p^{train},
\end{equation}
obtained through linear regression in order to give the best relation between the properties, $p^{train}$, and the compounds in the training set. The regularizer, $\lambda$ (multiplied with the unit matrix $I$), is usually small assuming that the noise in the training set is negligible. The non-linearity is given by the shape of the kernel~\cite{Rupp2015IJQC,Faber2018JCP}. When using the recently developed FCHL representation~\cite{Faber2018JCP}, atomistic Gaussian kernels are used as for example in Ref.~\cite{Faber2018JCP}. 
The standard FCHL model can only predict the energy of one electronic state at a time and thus, the KRR model has to be used in three versions for the three different electronic states studied here.

In order to find a model that is able to predict values for each electronic state at once, we also used extended representations that encode the quantum energy level (see above). An additional Gaussian kernel was used in this case and subsequently combined with the original kernel that maps a compound to its property. The final model was able to predict $N_s$ energetic values for one given representation with $N_s$ being the number of different electronic states:

\begin{equation}
    \label{equ:main_kernel_state}
    \boldsymbol{p}(M,S)=\sum_{k=1}^{N_m}\sum_{l=1}^{N_s}\alpha^l_k K_1(M,M_k) K_2(S,S_l).
\end{equation}
Here, $K_1$ and $K_2$ are different kernel matrices and $\mathbf{p}(M,S)$ is a vector of length $N_m \cdot N_s$, which can be recast as a matrix of size $N_m \times N_s$. S refers to the electronic state.\\

Assuming correct learning behaviour, a ML model must show decreasing out of sample error with increasing training set size, as has been shown by Vapnik and coworkers for KRR~\cite{Corinna1994} and by M\"uller and coworkers for NNs~\cite{Mueller1996NC}. 
To obtain such a learning behavior is easier to obtain with KRR compared to NNs because the number of hyperparameters is usually smaller for KRR than for NNs. The ease of use is thus a clear advantage of KRR~\cite{Faber2017JCTC,Faber2018JCP}.

\subsection{Multi-layer feed forward neural networks (NNs)}

In this study, simplistic multi-layer feed forward neural networks (NNs) were used. In contrast to high-dimensional NNs, which contain one NN for each atom type of a molecule and were developed with the idea that the energy of a molecule can be given as a sum of its atomic contributions~\cite{Behler2015IJQC}, only one NN is used for the complete molecule in this work. We selected this NN architecture since we treat only one molecule here and because the fragmentation capabilities of high-dimensional NNs seem to be inferior for excited-states than for the ground state \cite{Westermayr2019CS}.
Compared to KRR, our NNs possess more hyperparameters and, thus, are more difficult to optimize with respect to error convergence. However, due to their architecture, their benefit lies in their flexibility and possibility to relate a molecular structure to a many-state output. In principle, this many-state output can be obtained without additional encodings like the state kernel $K_2(S,S_l)$ in our KRR approach. Nevertheless, we sought to investigate the influence of encoding the respective quantum energy level for comparison to the KRR approach. Therefore, we also used state-encoding representations as detailed above.

To examine the impact of all properties treated within one ML model, we use a NN, abbreviated to NN-All in the following. As an input representing molecular geometries, we stay with the matrix of inverse distances and FCHL representation for comparison issues, but as an output, we employ a multidimensional vector including all properties -- energies, corresponding  forces, and permanent dipole moments as single values for each electronic state as well as transition dipole moments and NACs between each set of electronic states. Also here, two different approaches for the gradients are used - once they are included in the output vector as single values and once they are treated as derivatives from the ML potentials for energies. We want to note, that we did not test on a state encoding for this model, because some outputs can not be attributed to a single state, in contrast to the aforementioned models trained on energies. Permanent dipole moments are state-specific, while transition dipole moments and NACs arise from two different electronic states. In total, the number of output features is 138 for the model that is directly trained on forces and 84 for the model that treats forces as derivatives.

All NN models use the numpy~\cite{Walt2011CSE} and theano~\cite{TDT2016a} distribution implemented in python. To find optimal hyperparameters of the models to represent the relation between a molecular geometry and its multi-dimensional output, random grid search of different sets of hyperparameters was carried out, see Ref. \cite{Westermayr2019CS} for details. In all cases, the stochastic gradient descent optimization algorithm Adam~\cite{Adam2014} (adaptive moment estimation) was applied and the learning rate was annealed during training with an early-stopping mechanism that took care of overfitting. When forces, $F$, are treated as potential derivatives, they are included in the loss-function, $L_{p,F}$, of the mean squared errors between energies or all properties, $p$, predicted by NNs, $p_k^{NN}$, and the reference value, $p_k^{QC}$:

\begin{eqnarray}
\centering
\label{eq:2}
L_{p,F} &=& \frac{1}{N_m} \sum_k^{N_m} (p_k^{NN}-p_k^{QC})^2 \\
&&+\frac{1}{N_m}\sum_k^{N_m}\frac{1}{3N_a}\sum_{\alpha}^{3N_a}(F_{k_{\alpha}}^{NN}-F_{k_{\alpha}}^{QC})^2. \nonumber
\end{eqnarray}

$F_{k_{\alpha}}^{NN}$ are the values of forces predicted with NNs and $F_{k_{\alpha}}^{QC}$ are corresponding reference values, where $\alpha$ runs over all atoms, $N_a$. 

We included the mean squared errors of all properties into the loss function of the NN-All model according to
\begin{equation}
    L_{p}=\frac{1}{N_p}\sum^{N_p} \frac{1}{N_m} \sum_k^{N_m}(p_k^{NN}-p_k^{QC})^2
\end{equation}
with $N_p$ being the number of properties. Since all properties, except for gradients, are scaled according to their mean and standard deviation, each value is weighted equally in the loss function. Gradients were also tested as derivatives of NN potentials for energies in case of the NN-All model, therefore they are added to the loss function as given in equation \ref{eq:2}.

\section{Results}
In the following, we assess the performance of the different ML approaches and each method combination is abbreviated as "ML model/representation". 
 
\subsection{ML nonadiabatic molecular dynamics}

\begin{figure}[h]
    \centering
    \includegraphics[scale=0.25]{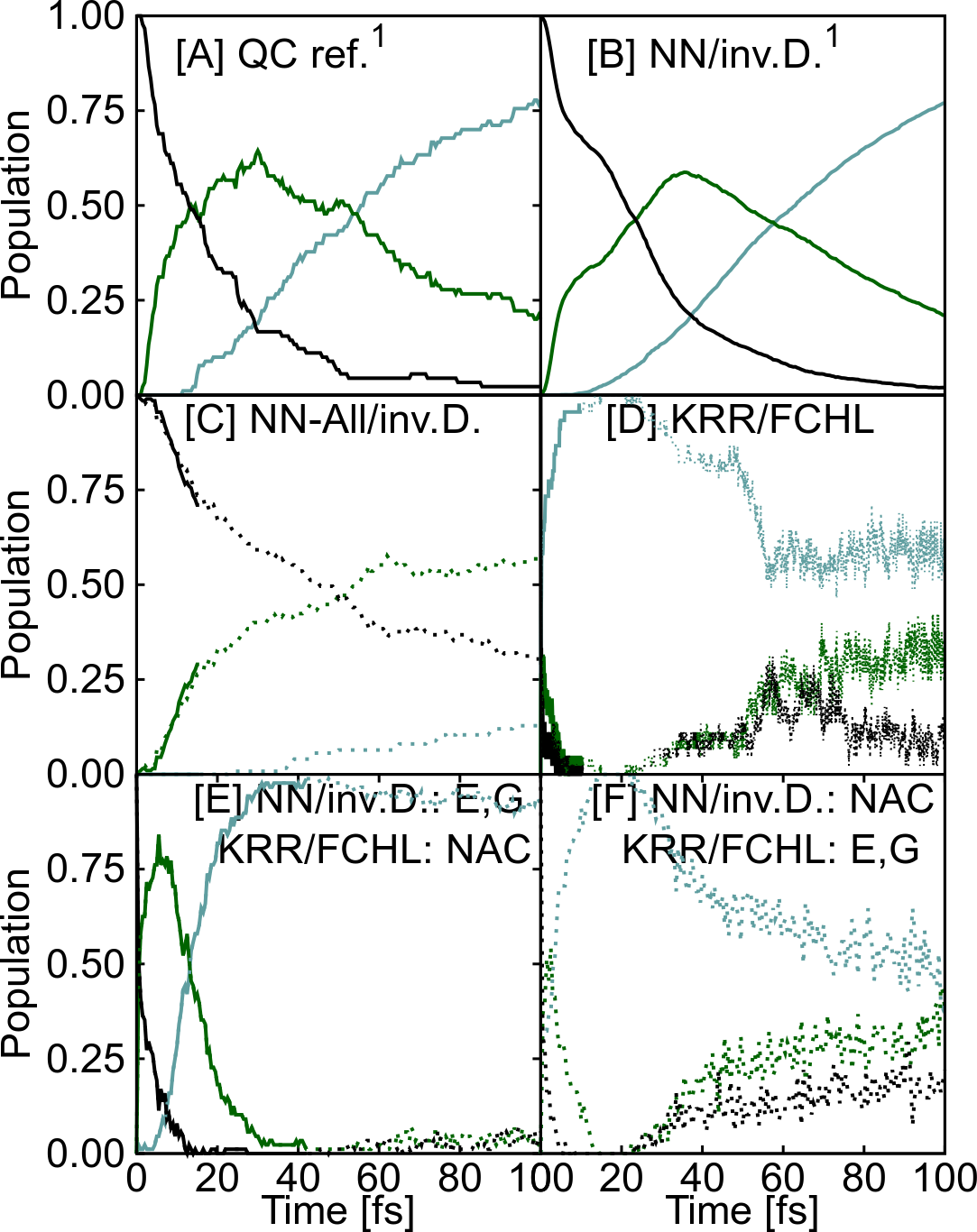}
    \caption{\label{main_dynamics}Populations of three singlet states as a function of time obtained from surface-hopping molecular dynamics simulations using six different methods.
    Dynamics according to [A] the quantum chemistry reference$^1$ (MR-CISD(6,4)/aug-cc-pVDZ, labeled as QC), [B] NN/inv.D.$^1$, [C] NN-All/inv.D., [D] KRR/FCHL and mixed ML models (energies (E) and gradients (G) from NN/inv.D. and NACs from KRR/FCHL in panel [E] and vice versa in [F]) are obtained after excitation of the methylenimmonium cation to the second excited singlet state (black lines). Green and turquoise lines correspond to first excited and ground state. Dotted lines are populations that are considered to be wrong with respect to the quantum chemical reference in panel (A) or large energy fluctuations ($^1$ taken from Ref.~\cite{Westermayr2019CS}).}
\end{figure}
The purpose of our ML models trained on energies, forces, and NACs is to successfully reproduce surface-hopping molecular dynamics simulations of the reference method. The populations obtained from surface-hopping molecular dynamics simulations with chosen ML models that take gradients as derivatives from ML energy potentials are given in Fig.~\ref{main_dynamics}. The reference dynamics in Fig.~\ref{main_dynamics} (A) results from initially 200 trajectories (thereof 90 trajectories for final use) and is taken from Ref.~\cite{Westermayr2019CS}, as is the population scheme in Fig.~\ref{main_dynamics} (B), that results from a NN/inv.D. model trained on 4000 training points and 3846 trajectories. It demonstrates one of the goals of ML dynamics -- to achieve better statistics at lower computational costs. Panels (C) to (H) of Fig.~\ref{main_dynamics} are obtained from the same 200 initial conditions as panel (A). If necessary, trajectories with large total energy fluctuations or jumps larger than 1 eV in the potential or kinetic energy are sorted out with tools of the SHARC~\cite{sharc-md2} program. Each dynamics simulation is carried out for 100 fs. As can be seen, the agreement between quantum chemistry (plot (A)) and NN/inv.D. (plot (B)) is fairly good. All the other models do not give an equivalent agreement to the reference method. 

Panel (C) shows results from NN-All/inv.D. calculations, which agree with the reference scheme up to around 10 fs. During this time, the population is transferred from the $S_2$ state to the $S_1$ state. Afterwards, there are fewer hops than there are expected to take place and after the simulation time of 100 fs, only a small fraction of the population is in the $S_0$ state, whereas most of the population should be transferred to the $S_0$ state according to the reference method. The main cause for this behavior are wrong hopping probabilities, which result from inaccurate NAC vectors. If the NACs are too small, too few hops take place, which is the case with this model. Accordingly, dotted lines are used instead of solid lines when the dynamics results are judged as unreliable.

The population scheme resulting from dynamics with the KRR/FCHL model is given in panel (D). As can be seen, within the first 10 fs, all of the population is transferred to the electronic ground state, $S_0$. After that, there are hops from lower lying states to higher energetic states. Those hops, especially in cases of large potential energy gaps between states, are considered to be implausible and the trajectories are not reliable anymore. Furthermore, the molecule atomizes during the course of the simulation, which is not the case in the quantum chemistry reference dynamics, indicating wrong ML dynamics and not converged ML potentials. The premature population transfer leads the molecule to regions of the conformational subspace, that are not visited with the reference method and are also not considered in the training set. Surprisingly, cuts through the potentials look very accurate when comparing KRR/FCHL with the quantum chemistry reference method or NN/inv.D., as will be discussed below (see Fig.~\ref{main_PEC}). Only close to critical regions of the PES, the NNs fit the quantum chemistry curves slightly better.

In order to find out if those small differences lead to such wrong dynamics, we carry out molecular dynamics simulations with mixed models. We combine the energies and gradients from the NN/inv.D. model as used for panel (B) in Fig.~\ref{main_dynamics} with the NACs from the KRR/FCHL model as used for panel (D). The results are given in panel (E) in Fig.~\ref{main_dynamics}. As can be seen, the mixed model reproduces the trend of the population transfer, but faster than it is the case in the reference dynamics. After around 40 fs, the population stays in the $S_0$ state and only few hops take place. These results imply that the NNs learn NACs more accurately than KRR and, moreover, that the slightly more accurate PESs of NN/inv.D. are necessary to converge to correct dynamics.

In turn, we also investigate the populations of a mixed model that takes energies and gradients from KRR/FCHL and NACs from the NN/inv.D. model. The obtained results in panel (F) show that the population trend can hardly be reproduced and again, the relaxation back to the ground state takes place much faster than it should be according to the reference method. As for KRR/FCHL in panel (B), a re-population of higher-lying states occurs after 20 fs, which disagrees with the reference dynamics.
A comparison of all these findings implies that an accurate prediction of energies is more important for reproducing the dynamics than an accurate prediction of NACs. Nevertheless, it is intuitively clear that surface-hopping molecular dynamics requires all properties to be accurate enough: Having the correct potentials, but completely wrong NAC values would also result in the wrong dynamics. 

A reason for the less accurate potentials of KRR might be the size of the training set, since it automatically fixes the matrix size of the kernel, while the depth (number of hidden layers and nodes) and thus complexity of the NN can be chosen independently of the training-set size. Here, we chose a number of 6 hidden layers and 50 nodes per hidden layer for all NN models resulting in NNs with higher complexity than KRR models for the given training set size. Up to a certain extent until the MAE converges, predictions should get more accurate, the deeper the model is.
Thus KRR trained on a larger training set should result in a comparable population scheme as in panels (A) and (B). Due to the fact that our available training set has only a limited amount of phase corrected data points, we did not further test this assumption. However, the assumption is supported by literature, where some dynamics simulations have been successfully carried out with KRR, see Ref.s~\cite{Dral2018JPCL,Hu2018JPCL}. At least 1.000 points are necessary to reproduce the two-state dynamics of a 1-D model system~\cite{Dral2018JPCL} or 60.000 data points for a system of 30 degrees of freedom~\cite{Hu2018JPCL},  respectively. To compare to our system, we do not only treat two energetic states but three and have a system of 12 degrees of freedom.  

In the following, we will discuss each of the properties separately and analyze different ML models and representations via learning curves and scatter plots to validate the aforementioned hypotheses and show ways to improve on the accuracy of the ML models.

\subsection{Energies and gradients}
\subsection*{Learning curves}

\begin{figure*}[h]
\centering
\includegraphics[width=6.2in]{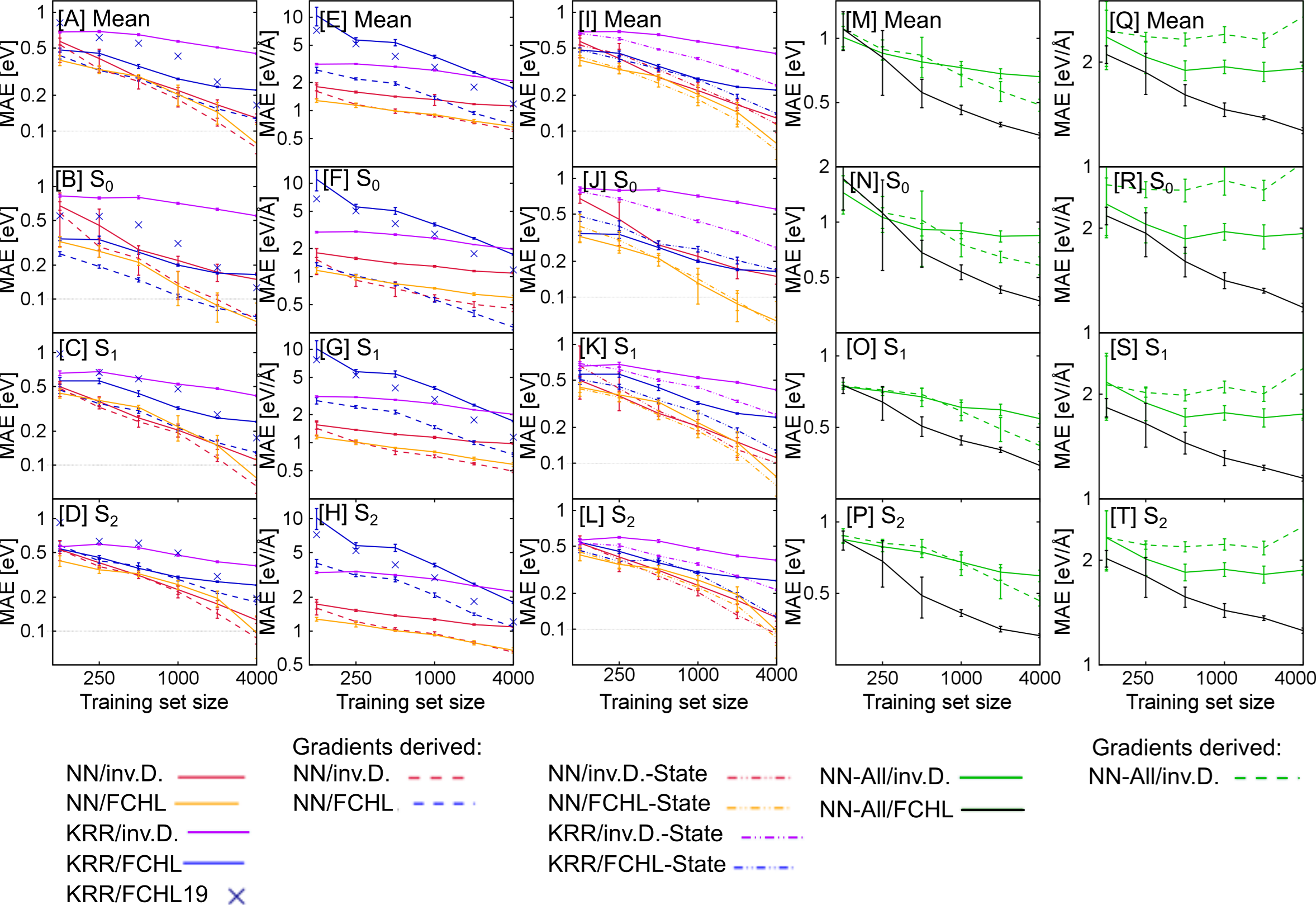}
\caption{\label{main_energy_grad_lc} Learning curves showing the mean absolute error (MAE) for the energies and gradients averaged over over all three singlet states in the first row and given separately for the states S$_0$, S$_1$, and S$_2$ in the second, third, and fourth row, respectively. FCHL and inv.D. denote the used representation. The first two columns show results from models that treat energies and gradients separately and together. The third column gives rise to models including a state representation and the last two columns show results from NN-All models, that treat all properties at once.}
\end{figure*}

\begin{figure*}[h]
\centering
\includegraphics[width=5in]{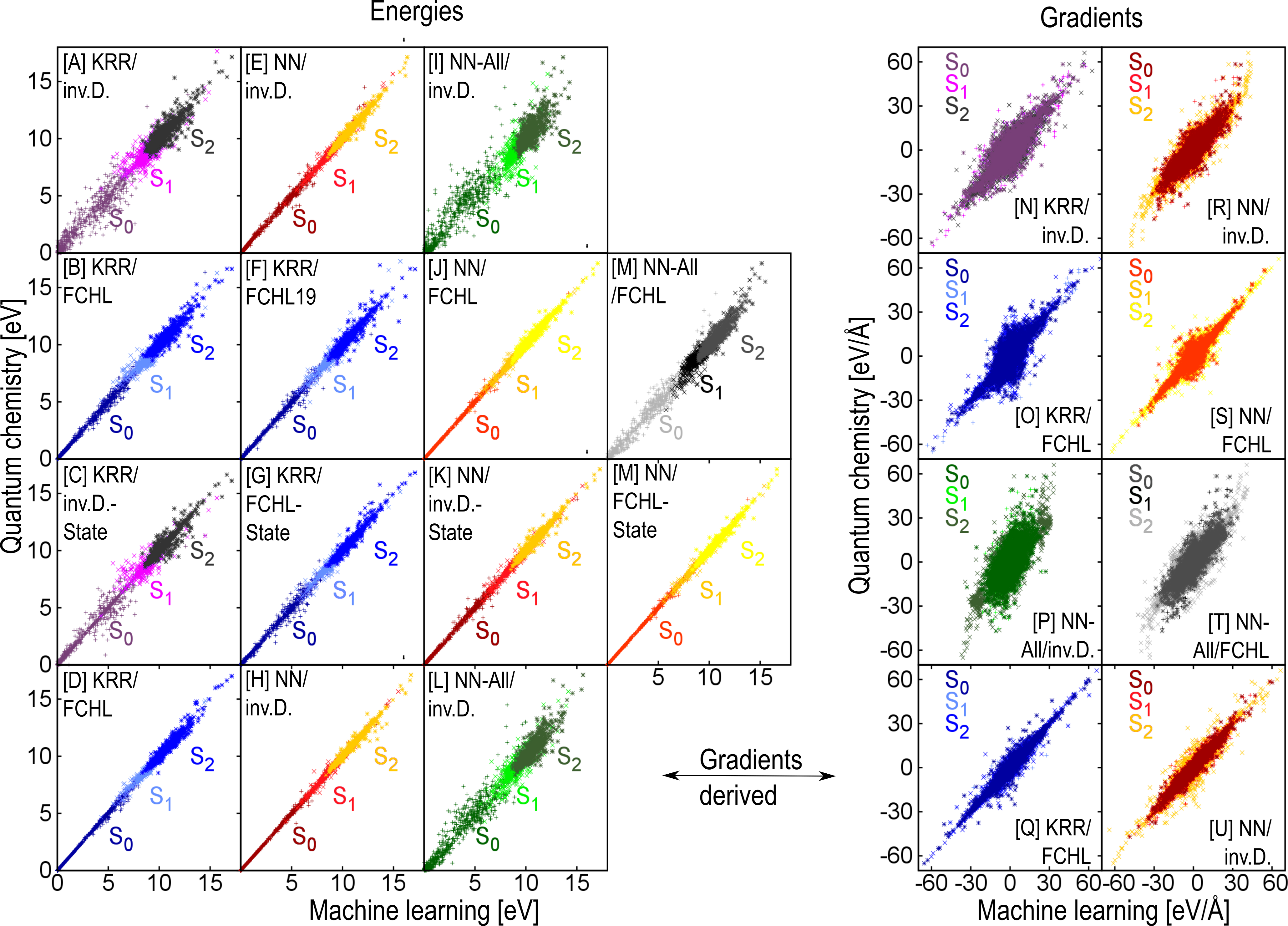}
\caption{\label{main_energy_scatter} Scatter plots for comparison of energies and gradients computed with different ML models and MR-CISD(6,4)/aug-cc-pVDZ. $S_0$, $S_1$, and $S_2$ denotes the electronic ground state, the first, and the second excited singlet state, respectively.}
\end{figure*}

First of all, we want to examine the learning of our ML models of energies and gradients. Learning curves for KRR and NNs -- showing the MAE of each electronic state as a function of training set size -- are given in Fig.~\ref{main_energy_grad_lc} with corresponding scatter plots given in Fig.~\ref{main_energy_scatter}. 

The first row of Fig.~\ref{main_energy_grad_lc} shows the mean MAE of all energetic states, whereas the second to the fourth row of Fig.~\ref{main_energy_grad_lc} give the learning curves of each electronic state separately. As already mentioned, the average of MAEs obtained from 10 different models is provided and the standard deviation is shown by means of error bars. The first, third and fourth column gives MAEs on energies and the second and fifth column on gradients. 

For dynamics simulations, it is crucial to obtain the forces as a derivative of the ML model in order to obey energy conservation. If gradients are learned as a quantity separate from energies, fluctuations in the total energy of a system occur within dynamics simulations~\cite{Li2015PRL,Chmiela2017SA,Dral2018JPCL,Hu2018JPCL}. Nevertheless, we investigated both variants for obtaining forces and study their influence for learning of energies and gradients. As it is visible from panels (A)-(H), learning the lowest state ($S_0$) is easiest for all models. The higher the electronic state becomes ($S_1$,$S_2$), the larger the MAEs become. All models that allow for derivation of gradients from ML models, allow to improve on the accuracy of energies as well as gradients -- which is the case for NN/inv.D. and KRR/FCHL (see the dashed lines in panels (A)-(H) compared to their respective solid counterparts).  

Comparing KRR with NN models, it is visible that NNs give lower MAEs than KRR models. Using the inv.D. representation for KRR results in rather high MAEs compared to the rest of the models. By improving the representation and using FCHL instead, the accuracy improves significantly. The same trend can be observed for NNs, but here, the inv.D. matrix already gives fair results that are comparable to KRR/FCHL results. NN/FCHL is comparable to NN/inv.D., that includes gradients in the training of energies. Hence, even more accurate models are expected for NN/FCHL with gradients treated as NN derivatives, which is not done here due to technical issues. Moreover, the number of hidden layers could be reduced from 6 to 4 by using the FCHL representation instead of the inv.D. matrix. Furthermore, the FCHL19 version is tested for energies and gradients using KRR and also shows improved results.

The reason for the lower prediction accuracy of KRR models is possibly because the NN models can train all energetic values at once, while KRR models are trained on each state separately. In order to test this assumption and improve the prediction accuracy of KRR models, we investigated the encoded state representation. The third column (panels (I)-(L)) shows results and compares different ML models with (dashed-dotted lines) and without (continuous lines) a state representation.  
As can be seen, the encoding of the quantum energy level improves the accuracy of all ML models, whereby this effect is significant for KRR models and small for NN models. The MAE of KRR/FCHL-State is comparable to NN/inv.D.-State. Still, the lowest MAE is obtained with NNs in combination with FCHL. The already high learning efficiency of NN/FCHL can not be improved significantly by encoding the quantum energy level. This observation indicates that the FCHL representation is already close to optimal, while the inv.D. representation employed here is not. Also the learning efficiency, that is dictated by the slope of the learning curves, is best for models using the FCHL representation. The highest learning efficiency can be achieved with the KRR/FCHL model, showing the steepest slopes especially for gradients, indicating that KRR outperforms NNs for larger training set sizes. By using the state representation for KRR, the kernel matrix size increases from $N_m \times N_m$ for KRR/inv.D. and KRR/FCHL to $N_s\cdot N_m \times N_s \cdot N_m$ for KRR/inv.D.-State and KRR/FCHL-State. In case of high memory consumption, this size can be reduced without a major loss in accuracy by mapping only a subset of molecules to the complete dataset for training. This makes the training process a lot more efficient.

The last two columns give results from NN-All models. One could assume that training on all properties at once would be beneficial since hidden patterns in the different data sets could lead to synergistic effects and an improved learning. However, as already observed from the dynamics simulations and as can be seen now in Fig.~\ref{main_energy_grad_lc} (N)-(T), a restriction of the training to only energies gives lower MAEs for energies than training on all properties at once (compare also scatter plots in Fig. \ref{main_energy_scatter}). In case of the gradients, panels (Q)-(T) show that treating all properties together and learning the gradients directly is more beneficial than treating them as derivatives and can even result in learning curves with positive slopes for the latter case. Nevertheless, the treatment of gradients improves the learning of energies and overall, the learning curve obtained as a sum of all properties shows a negative slope. This is shown in more details in the SI.

However, as pointed out above, for dynamics simulations gradients should be considered as potential derivatives, which does not lead to sufficiently accurate NN-All/inv.D. models for dynamics simulations. The MAE of energies is approximately 5 times larger and of gradients twice as large as the MAEs obtained with the NN/inv.D. model trained solely on energies and gradients, explaining the wrong results of the excited-state dynamics. This trend is also clearly visible in the scatter plots for gradients, see Fig \ref{main_energy_scatter} panels (N)-(U) and gives rise to the assumption, that a NN of equal depth, that is trained on only one property, is superior to a NN that is trained on more properties when considering the properties separately. Moreover, a deeper NN architecture did not result in lower MAEs or steeper learning curves. More advanced multi-task learning approaches~\cite{Ruder2017arXiv} were not tested for time reasons. 
Again, the FCHL representation gives higher prediction accuracy than the inv.D. matrix. Nevertheless, the NN-All/FCHL model is less accurate than NN/FCHL trained only on energies or gradients. Summarizing the NN findings, training on all properties that are rather different in nature thus has a negative effect on the learning of the separate properties. 

Those results can be further manifested with scatter plots given in Fig.~\ref{main_energy_scatter} and summarized as follows: It is clearly visible that (i) NNs (red plots) result in more accurate models for energies and gradients than KRR (blue and purple plots) using a limited training set of 4000 data points. (ii) As expected, the FCHL representation (second panel line) leads to higher prediction accuracy than the inv.D. representation (first panel line) and also results in steeper learning curves, especially for KRR. (iii) Explicitly encoding the electronic state (third panel line for energies) is beneficial for KRR models compared to single-state KRR models (first two rows) and (iv) treating all properties at once that are dissimilar in nature (greenish and grey plots) is disadvantageous. (v) Finally, treating gradients as derivatives of NN potentials and as response properties for KRR (last panel line in Fig.~\ref{main_energy_scatter}) leads to more accurate energies and gradients.

\subsection*{Potential energy scans}

 \begin{figure}[h]
    \centering
    \includegraphics[scale=0.25]{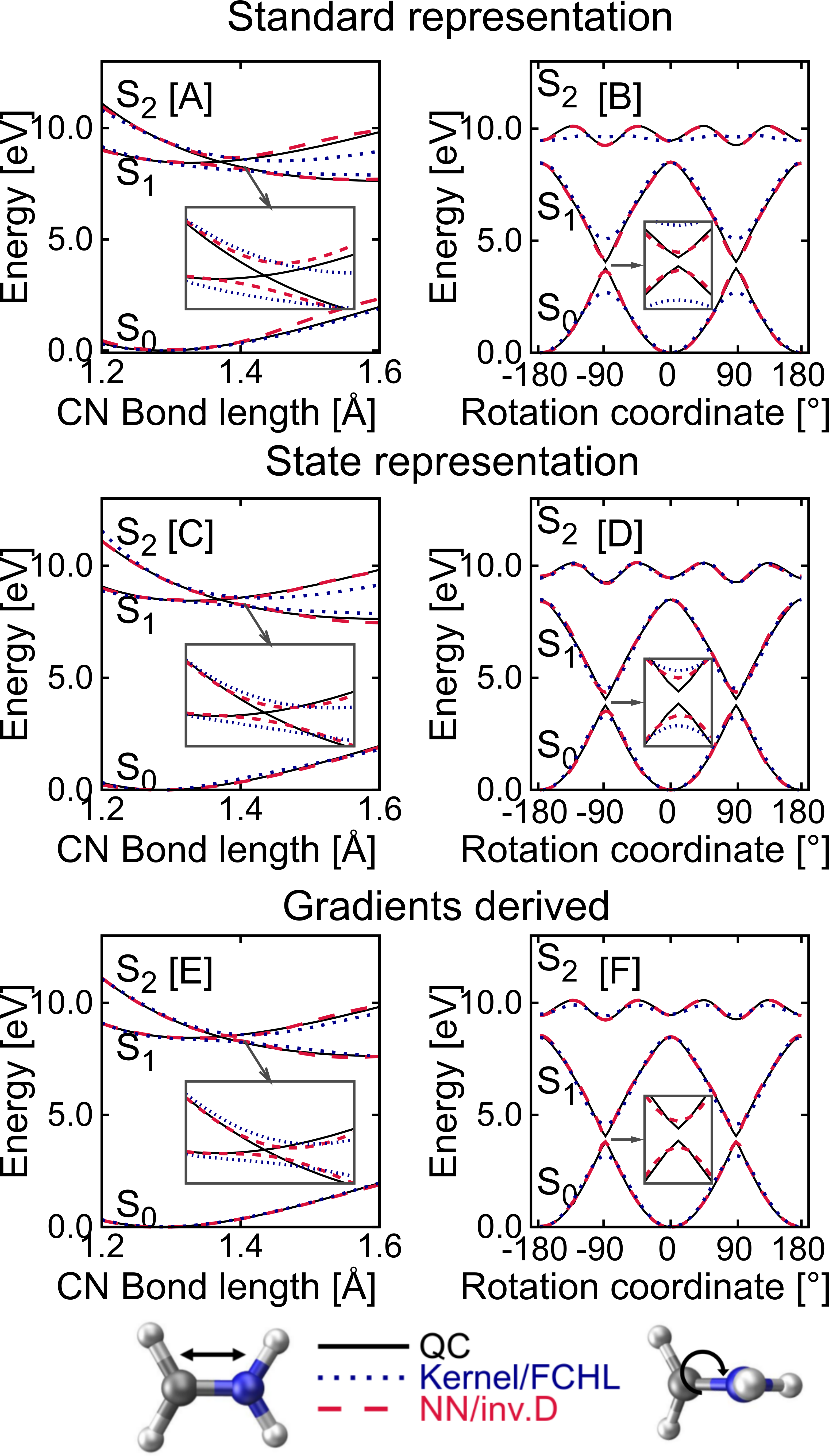}
    \caption{\label{main_PEC}Scans along the CN-bond elongation of the methylenimmonium cation (left panels) and the rotation along the dihedral angle (right panels) computed with quantum chemistry (QC, black continuous lines), KRR (blue dotted lines), and NNs (red dashed lines). ML models that were trained only on energies without (first row) and with (second row) a representation for the electronic state  are compared to ML models, that treat energies and forces together (third row).
    }
\end{figure}
 In addition to the learning curves and scatter plots, we investigate potential energy curves along two reaction coordinates that include critical regions important for photodynamics simulations. Results are given in Fig~\ref{main_PEC}. The left plots show the scan along the elongation of the bond between the carbon atom and the nitrogen atom of the methylenimmonium cation, whereas the right plots show the scan along the dihedral angle of the molecule. The first reaction coordinate leads to a $S_2/S_1$ CI and the second to a $S_1/S_0$ CI. Both reaction coordinates are thus crucial for excited-state molecular dynamics. The curves in Fig.~\ref{main_PEC}, obtained as cuts along the aforementioned coordinates, show the quantum chemistry reference method by black continuous lines, the forecasts of KRR/FCHL models by blue dotted lines and the ones of NN/inv.D. models by red dashed lines. We focus on those two models, because both can treat gradients as derivatives, which is crucial for dynamics simulations. Fig.~\ref{main_PEC} (A) and (B) represent results obtained without a state representation ("standard representation"), where energies are trained without forces. The second row (panels (C) and (D)) gives results obtained when the quantum energy level is encoded ("state representation") and the last row (panels (E) and (F)) illustrates results, that are obtained with ML models that treat gradients as derivatives of the ML models for the corresponding energies. 
\\
As can be seen, all plots generally show fair agreement of ML models to the quantum chemistry reference method. 
Taking a look at the first row, the inset of panel (A) visualizes that the KRR/FCHL curves are comparable to the reference data. The same is true for the NN/inv.D curves. 
Thus, both models can be seen as comparably accurate. 
Considering the ML models that include a representation for the quantum energy level in the second row, NN/inv.D. curves are slightly closer to the quantum chemistry reference than KRR/FCHL curves are. This observation holds for both reaction coordinates shown in panel (C) and (D). Overall, the agreement is better than it is without encoding the quantum energy level. 
Last, results from models that include gradients in training of energies are given in panels (E) and (F). As can be seen, NN/inv.D. is very close to the reference, even at the avoided crossings between the two singlet states. The potential energy curves obtained from KRR/FCHL do not show much improvement for energies compared to previous representations. Since the ML potentials have to be differentiable, they are smoother than the potentials obtained with quantum chemistry. Taking into account that gradients should be derived from ML models for dynamics simulations, the NN/inv.D. model that includes forces into training of energies is most promising for excited-state molecular dynamics simulations. Although the PECs plotted in Fig.~\ref{main_PEC} (E) and (F) yield similarly accurate results of ML models that treat gradients as derivatives, the dynamics simulations differ a lot from each other. These results support the hypothesis that the slightly more accurate ML potentials for energies and gradients are necessary to reproduce excited-state dynamics correctly, see Fig.~\ref{main_dynamics}.

\subsection{Nonadiabatic couplings (NACs)}
Another important property that should be considered for excited-state molecular dynamics simulations is the NAC vector between different states of same spin multiplicity, in this case singlet states. We therefore investigate the ability of our ML models to predict NAC values. On the one hand, we learn the NACs directly and on the other hand, we try to improve their prediction accuracy by scaling them with the energy gap of the two respectively coupled states (see equation~\ref{eq:naclearn}). 

\begin{figure}[h]
    \centering
    \includegraphics[scale=0.32]{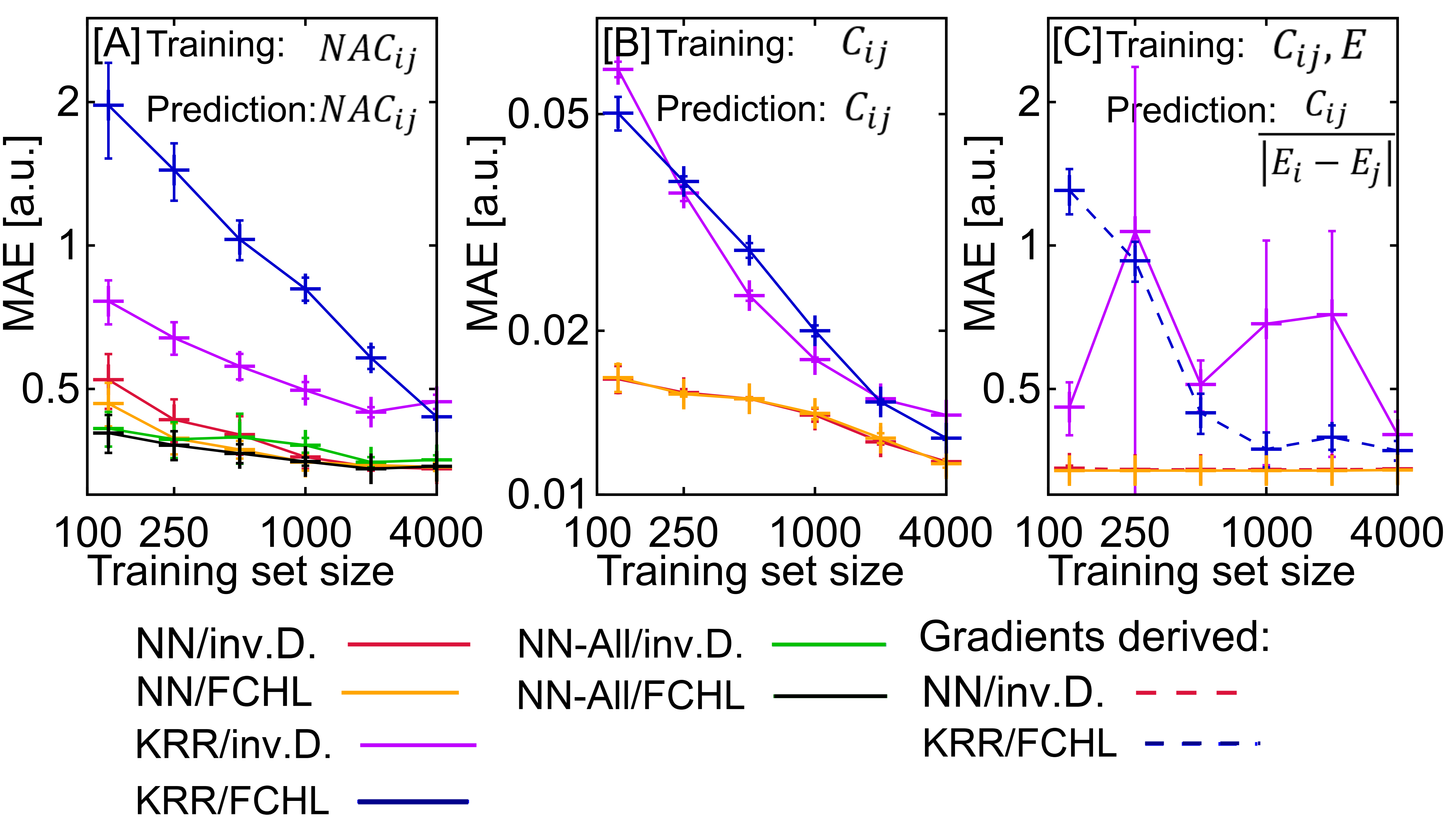}
    \caption{Learning curves showing the MAE of phase corrected NACs as a function of training set size for different ML models. Panel (A) shows learning curves of $NAC_{ij}$ (equation \ref{eq:3}), panel (B) of $C_{ij}$ (equation \ref{eq:naclearn}) and panel (C) of $NAC{ij}$ obtained from ML models trained on $C_{ij}$ in combinations with the best corresponding models for energies (equation \ref{eq:nacpredict}}.
    \label{main_nac_lc}
\end{figure}

Learning curves are given in Fig.~\ref{main_nac_lc}, where panel (A) shows the performance of ML models to predict NACs as they are obtained from quantum chemistry calculations after phase correction. NNs exhibit a rather flat learning curve, whereas KRR shows a steeper learning curve, hence more efficient learning. However, the NNs start already at a rather low error such that they still outperform KRR also at a large training set size. Again, the FCHL representation clearly outperforms the inv.D. matrix for KRR. For NNs, no clear advantage of one representation over the other can be observed. Remarkably, NN-All/inv.D. and NN-All/FCHL perform similarly as well. NN-All/inv.D. is slightly less accurate than NN/inv.D. and NN-All/FCHL performs simliar to NN/FCHL.
Panel (B) shows the results of the second approach, the training and prediction of $C_{ij}$ (see equation~\ref{eq:naclearn}), that includes the information about the coupled PESs. As mentioned before, multiplication of $NAC_{ij}$ values with the energy gap between the corresponding states should lead to smoother NAC curves and thus better learning. The mean of the absolute values of NACs directly obtained from QC is about 22 times larger than the mean of the absolute values of $C$s. Thus the MAE should also be at least a factor 22 lower for predicted $C_{ij}$ values.
As can be seen from the learning curves, this is the case, but the MAE decreases by an even larger factor -- that is about 34 for both, KRR/inv.D. and KRR/FCHL models, and approximately 30 for NN/inv.D. as well as NN/FCHL. The learning efficiency of KRR/inv.D. increases as well, while the learning efficiency of NN models does not improve. Nevertheless, the latter models still show learning behaviour for larger training set sizes in comparison to the the NN learning curves in panel (A), that flattens towards larger training set sizes.  Furthermore, the standard deviations between the models are smaller, hence more reliable ML models are obtained.

To assess whether the ML PESs are accurate enough to rescale from the $C_{ij}$ to the actual $NAC_{ij}$ values, the best ML models for energies are combined with the ML models trained on $C_{ij}$ (panel (B)). 
As can be seen from the learning curves, the MAE converges towards the same values as in panel (A) in case of NNs and decreases for KRR/inv.D. and KRR/FCHL. Moreover, the MAE of NNs stays approximately the same, which indicates that the quality of NACs from NNs can not be improved using this approach. 
Another method to improve the NACs could be the operator formalism~\cite{Christensen2019arXiv}, that was already applied for dipole moments and gradients and could be adapted for NACs as well. As an outlook, we expect that the learning efficiency and predictions accuracy of NACs by treating them as so-called response properties can further be improved.

Another important finding is that including the energy gap for training and prediction of NACs results not only in higher prediction accuracy for KRR, but also less stable predictions. This statement can be verified by looking at the larger standard deviations. Especially in the case of KRR/inv.D., the learning curve is not linear, but shows rather random jumps in the MAEs in combination with very large uncertainties. This result can be explained by taking a look at Fig.~\ref{main_energy_grad_lc} (A) and ~\ref{main_energy_scatter} (A), which shows the low prediction accuracy of this model for energies. Hence, the inaccurate energy gaps are used to correct the magnitude of the learned NAC values, resulting in instable ML models for NACs. Also for KRR/FCHL, the standard deviations enlarge.
We conclude that the approach of scaling NACs with the energy gaps is unfavorable in the present case.

\subsection{Dipole Moments}

\begin{figure}[h]
    \centering
    \includegraphics[scale=0.65]{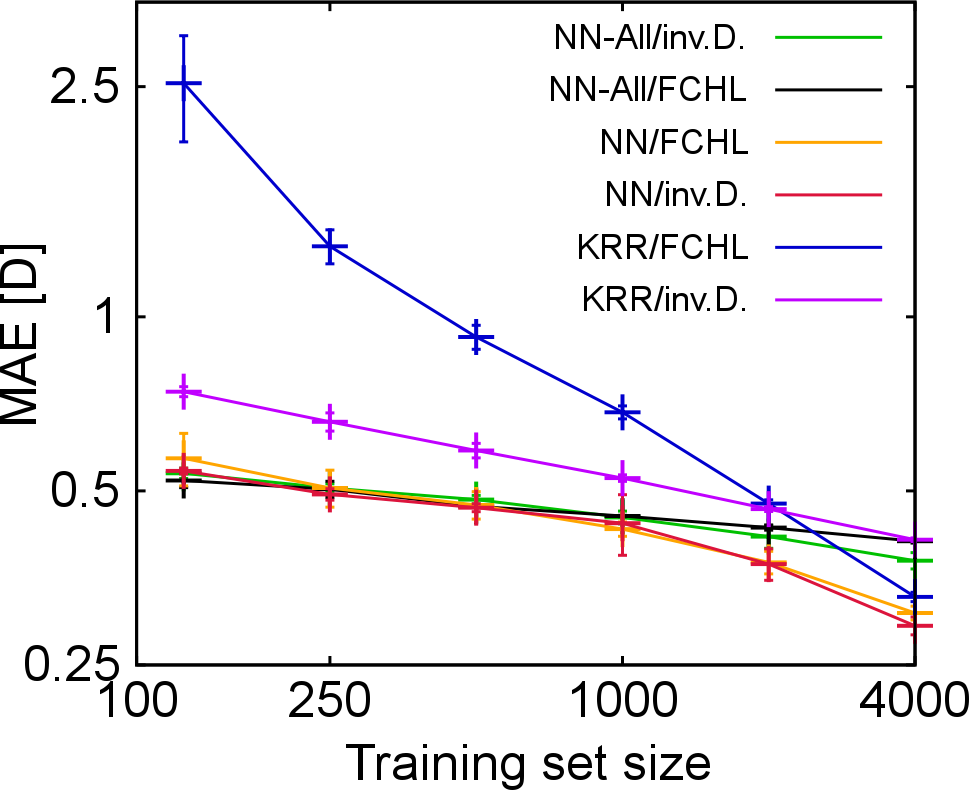}
    \caption{\label{main_dipole_lc}Learning curves that show the MAE of  phase corrected (transition) dipole moments as a function of training set size for different ML models.
    }
\end{figure}

Dipole moments are important properties of molecules and can be used to compute IR spectra, see for example Ref.~\cite{Harris1989}. ML models to accurately predict dipole moments already exist~\cite{Ramakrishnan2014SD,Artrith2011PRB,Huang2016JCP,Gastegger2017CS,Yao2018Cs,Schuett2018JCP,Nebgen2018JCTC,Sifain2018JPCL,Schuett2019,Schuett2019JCTC,Christensen2019JCP} with different approaches, such as charge models~\cite{Gastegger2017CS,Sifain2018JPCL} or the response formalism~\cite{Christensen2019JCP}, being applied. Transition dipole moments are employed in the calculation of pump-probe schemes, static-field interactions, or time-resolved spectra, see for example Refs.~\cite{Marquetand2004JCP,Bonafe2018JPCL}.
Here, we train on the permanent and phase corrected transition dipole moments of three singlet states and compare different representations and regressors. Fig~\ref{main_dipole_lc} shows the learning curves of the used models.
 As in the case of NACs, the highest learning efficiency is obtained with KRR/FCHL, indicating again that KRR/FCHL might outperform NNs using more data points. The learning efficiency of KRR/inv.D. is comparable to the one of NN/inv.D. and NN/FCHL, whereas the NN-All models show less efficient learning. In contrast to the NAC values, where KRR/FCHL is worse than NN-All/FCHL, the NN-All models yield less accurate dipole moments than KRR/FCHL. The lowest MAE, hence most accurate predictions, are obtained with NN models. Using the FCHL representation instead of the inv.D. matrix does not lead to improved prediction accuracy in this case. By using the operator formalism~\cite{Christensen2019JCP}, we again assume that the prediction accuracy can be further optimized. 

\subsection{Principal component analysis of ML models}
\begin{figure}
    \centering
    \includegraphics[scale=0.35]{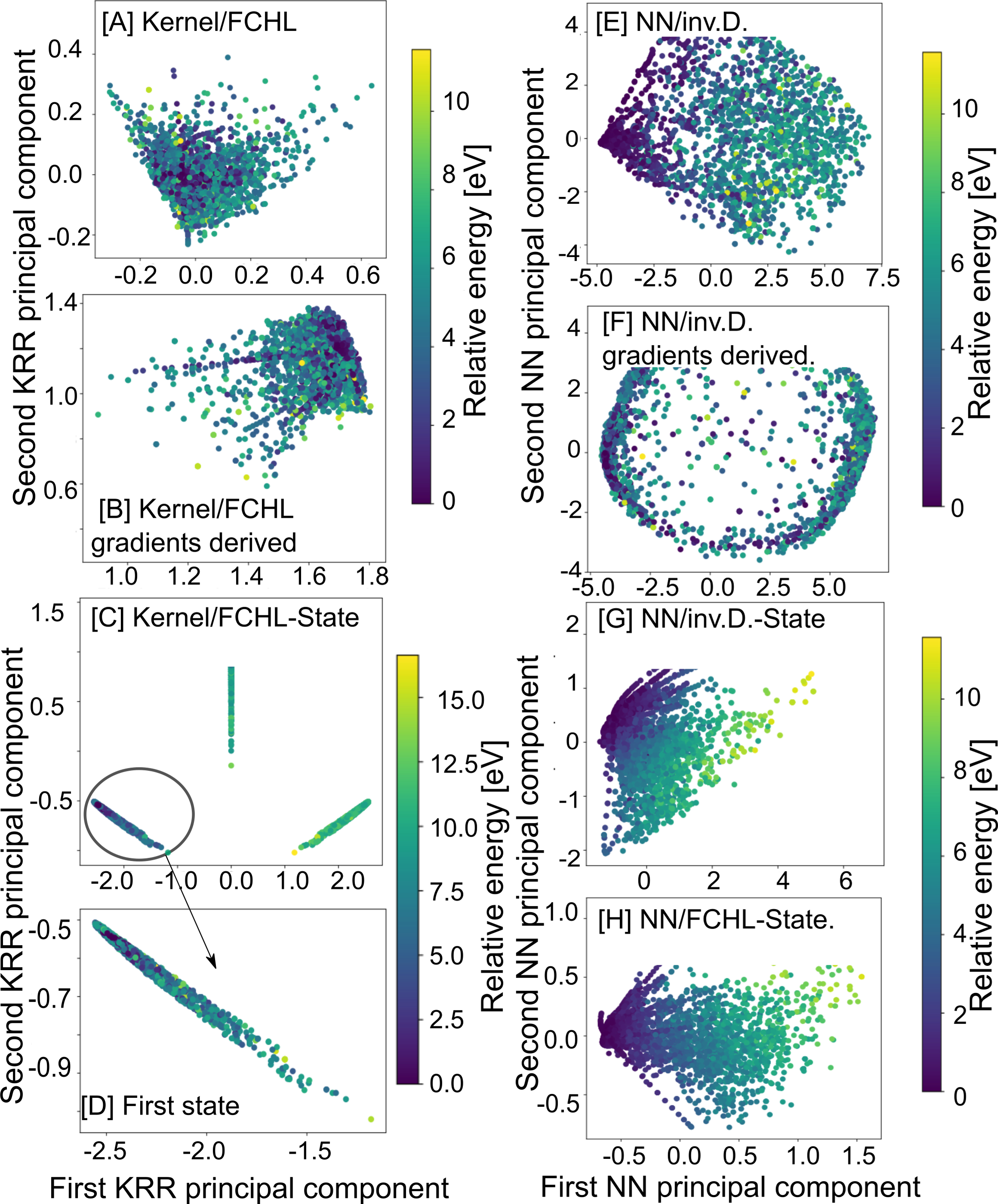}
    \caption{\label{main_PCA}Principal component analysis of the kernel matrices of different models and the matrices obtained from the last NN-layers. Each vector in the NN-matrix corresponds the last layer of one input. The left side gives results for KRR, whereas the right side shows results obtained from NNs.}
\end{figure}

In order to obtain further insights, we carried out principal component analyses of the models used for surface-hopping molecular dynamics simulations and compare to the same models including a state representation. Graphs showing the first principal component plotted against the second principal component of the respective ML model are given in Fig.~\ref{main_PCA}. On the left side, results from KRR are illustrated. Remarkable is that the KRR/FCHL-State model, that includes a representation to encode the quantum energy level (Fig.s~\ref{main_PCA} (C) and (D)), shows a clear ordering of the data corresponding to different electronic states. Moreover, within one state, a better ordering can be obtained than for KRR/FCHL with and without the inclusion of gradients (panels (A) and (B)) but without a representation for the electronic states. 
Similarly, NNs, visualized in the right plots of Fig.~\ref{main_PCA} in panels (E)-(H), show an improved ordering of data, when a representation for the electronic state is included. Nevertheless, also the model without the state representation (E) shows some kind of ordering of data according to energies, whereas the model that includes gradients as derivatives for learning of energies is less organized (F). These results suggest that the representation is much more important for KRR than it is for NNs, which are already able to order data points to some extent with the simplest representation used. This finding indicates once again the higher flexibility of NNs compared to KRR.
\\

\section{Conclusion}
In this paper, we compare two frequently used ML regressors, namely KRR and NNs, for their application in excited-state molecular dynamics of CH$_2$NH$_2^+$. We investigate their role in combination with different representations of the molecular structure for the prediction of energies of the ground state as well as excited states, corresponding forces, NACs between different states, and (transition) dipole moments. All ML models are able to learn the relation between a quantum chemical property and the molecular structure, when the properties are treated separately from each other. Learning all properties at once leads to significantly worse results and the learning of single properties can even be impeded when the cost function includes all properties at once.
\\
It is shown that the FCHL representation used as a representation for KRR and NNs is in most cases superior to the matrix of inverse distances. The results can also be improved by encoding the quantum energy level in the representation. In this way, we obtain a KRR approach with multiple outputs. The state encoding was shown for three electronic singlet states for KRR as well as NNs. In both cases, the modification of the representation is necessarily accompanied by an enlargement of the ML model, a larger kernel matrix in the case of KRR, and a larger input layer in the case of NNs. Principal component analyses further show that an enhanced ordering of data points is obtained by using a state-representation in addition to the molecular representation.
\\
Carrying out excited-state molecular dynamics simulations shows the large influence of the small error differences between the two ML regressors. While NNs can reproduce dynamics of the {\emph ab initio} reference method, as suggested from learning curves, KRR-based models need more data points in the training set to converge to similar dynamics outcomes. However, the slopes of KRR/FCHL learning curves are steeper, indicating that KRR might outperform NNs for large training sets, albeit with significant computer memory requirements. The NAC vectors pose a real challenge for ML models due to their peaked nature. While their accuracy can be improved for KRR/FCHL by including the energy gap of the coupled pair of states, this approach does not allow for more accurate NAC vectors due to deteriorating effects of the errors in the energy gaps.
\\
Summarizing these findings, it is not obvious that one ML model outperforms the other. It  rather depends on the various options which model is preferable. While KRR for example might show better performance for larger training set sizes and is generally easier to optimize with respect to hyperparameters,  NNs give good results already for small training set sizes, but hyperparameter optimization is tedious.
We hence recommend to use KRR for first explorative runs and switch to NNs for final production runs of excited-state dynamics simulations.
Future applications might use the concept of \textit{wide and deep learning}~\cite{Cheng2016arXiv} in the sense that different ML models can be applied within one applications to combine their distinct benefits.

\section*{Conflicts of interest}
There are no conflicts of interest to declare.
\section*{Acknowledgements}
This work was financially supported by the Austrian Science Fund, W 1232 (MolTag), the uni:docs program of the University of Vienna, (J.W.) and the University of Basel. The computational results presented have been achieved in part using the Vienna Scientific Cluster (VSC). We additionally want to thank Michael Gastegger from TU Berlin and Sebastian Mai from University of Vienna for discussions concerning nonadiabatic couplings and suggestions to improve on their prediction.
O.A.v.L. acknowledges funding from the Swiss National Science foundation (No.~PP00P2\_138932 and 407540\_167186 NFP 75 Big Data) and from the European Research Council (ERC-CoG grant QML). This work was partly supported by the NCCR MARVEL, funded by the Swiss National Science Foundation.
\section*{References}

\begin{thebibliography}{10}
\expandafter\ifx\csname url\endcsname\relax
\expandafter\ifx\csname urlprefix\endcsname\relax\def\urlprefix{URL }\fi
\providecommand{\eprint}[2][]{\url{#2}}

\bibitem{Vass2007ANAS}
Vass I, Cser K and Cheregi O 2007 {\em Ann. NY Acad. Sci.\/} {\bf 1113}
  114--122

\bibitem{Schoenlein1991S}
Sch\"{o}nlein R, Peteanu L, Mathies R and Shank C 1991 {\em Science\/} {\bf
  254} 412--415

\bibitem{Garavelli1999JACS}
Garavelli M, Negri F and Olivucci M 1999 {\em J. Am. Chem. Soc.\/} {\bf 121}
  1023--1029

\bibitem{Schreier2007S}
Schreier W~J, Schrader T~E, Koller F~O, Gilch P, Crespo-Hern\'{a}ndez C~E,
  Swaminathan V~N, Charell T, Zinth W and Kohler B 2007 {\em Science\/} {\bf
  315} 625--629

\bibitem{Rauer2016JACS}
Rauer C, Nogueira J~J, Marquetand P and Gonz\'alez L 2016 {\em J. Am. Chem.
  Soc.\/} {\bf 138} 15911--15916

\bibitem{Ahmad2016IJP}
Ahmad I, Ahmed S, Anwar Z, Sheraz M~A and Sikorski M 2016 {\em Int. J.
  Photoenergy\/} {\bf 2016} 1--19

\bibitem{Haese2016CS}
H\"ase F, Valleau S, Pyzer-Knapp E and Aspuru-Guzik A 2016 {\em Chem. Sci.\/}
  {\bf 7}(8) 5139--5147

\bibitem{Butler2018N}
Butler K~T, Davies D~W, Cartwright H, Isayev O and Walsh A 2018 {\em Nature\/}
  {\bf 559} 547--555 ISSN 1476-4687

\bibitem{Liu2015JPCL}
Liu J and Prezhdo O~V 2015 {\em J. Phys. Chem. Lett.\/} {\bf 6} 4463--4469

\bibitem{Domcke2013NC}
Domcke W and Sobolewski A~L 2013 {\em Nat. Chem.\/} {\bf 5} 257--258

\bibitem{Liu2017SR}
Liu F, Du L, Zhang D and Gao J 2017 {\em Sci. Rep.\/} {\bf 7} 1--12

\bibitem{Mai2018WCMS}
Mai S, Marquetand P and Gonz\'alez L 2018 {\em WIREs Comput. Mol. Sci.\/} {\bf
  8} e1370

\bibitem{Mai2016JPCL}
Mai S, Marquetand P and Gonz\'alez L 2016 {\em J. Phys. Chem. Lett.\/} {\bf 7}
  1978--1983

\bibitem{Doltsinis2006NIC}
Doltsinis N~L 2006 {\em Molecular Dynamics Beyond the Born-Oppenheimer
  Approximation: Mixed Quantum-Classical Approaches\/} ({\em NIC Series\/}
  vol~31) (John von Neuman Institut for Computing)

\bibitem{Subotnik2016ARPC}
Subotnik J~E, Jain A, Landry B, Petit A, Ouyang W and Bellonzi N 2016 {\em
  Annu. Rev. Phys. Chem.\/} {\bf 67} 387--417

\bibitem{Marquetand2017M}
Marquetand P, Nogueira J~J, Mai S, Plasser F and Gonz\'alez L 2017 {\em
  Molecules\/} {\bf 22}

\bibitem{Behler2015IJQC}
Behler J 2015 {\em Int. J. Quantum Chem.\/} {\bf 115} 1032--1050

\bibitem{Rupp2015IJQC}
Rupp M 2015 {\em Int. J. Quantum Chem.\/} {\bf 115} 1058--1073 ISSN 1097-461X

\bibitem{Gastegger2016JCP}
Gastegger M, Kauffmann C, Behler J and Marquetand P 2016 {\em J. Chem. Phys.\/}
  {\bf 144} 194110

\bibitem{Hansen2013JCTC}
Hansen K, Montavon G, Biegler F, Fazli S, Rupp M, Scheffler M, von Lilienfeld
  O~A, Tkatchenko A and M\"uller K~R 2013 {\em J. Chem. Theory Comput.\/} {\bf
  9} 3404--3419

\bibitem{Shen2018JCTC}
Shen L and Yang W 2018 {\em J. Chem. Theory Comput.\/} {\bf 14} 1442--1455

\bibitem{Wang2017PCCP}
Wang S, Yuan J, Li H and Chen M 2017 {\em Phys. Chem. Chem. Phys.\/} {\bf 19}
  19873--19880

\bibitem{Gastegger2017CS}
Gastegger M, Behler J and Marquetand P 2017 {\em Chem. Sci.\/} {\bf 8}(10)
  6924--6935

\bibitem{Li2015PRL}
Li Z, Kermode J~R and De~Vita A 2015 {\em Phys. Rev. Lett.\/} {\bf 114}(9)
  096405

\bibitem{Botu2017JPCC}
Botu V, Batra R, Chapman J and Ramprasad R 2017 {\em J. Phys. Chem. C\/} {\bf
  121} 511--522

\bibitem{Artrith2017PRB}
Artrith N, Urban A and Ceder G 2017 {\em Phys. Rev. B\/} {\bf 96}(1) 014112

\bibitem{Behler2017ACIE}
Behler J 2017 {\em Angew. Chem. Int. Edit.\/} {\bf 56} 12828--12840 ISSN
  1521-3773

\bibitem{Chmiela2017SA}
Chmiela S, Tkatchenko A, Sauceda H~E, Poltavsky I, Sch{\"u}tt K~T and
  M{\"u}ller K~R 2017 {\em Sci. Adv.\/} {\bf 3}

\bibitem{Behler2008PRB}
Behler J, Reuter K and Scheffler M 2008 {\em Phys. Rev. B\/} {\bf 77} 115421

\bibitem{Carbogno2010PRB}
Carbogno C, Behler J, Reuter K and Gro\ss{} A 2010 {\em Phys. Rev. B\/} {\bf
  81}(3) 035410

\bibitem{Dral2018JPCL}
Dral P~O, Barbatti M and Thiel W 2018 {\em J. Phys. Chem. Lett.\/} {\bf 9}
  5660--5663

\bibitem{Hu2018JPCL}
Hu D, Xie Y, Li X, Li L and Lan Z 2018 {\em J. Phys. Chem. Lett.\/} {\bf 9}
  2725--2732

\bibitem{Xie2018JCP}
Xie C, Zhu X, Yarkony D~R and Guo H 2018 {\em J. Chem. Phys.\/} {\bf 149}
  144107

\bibitem{Chen2018JPCL}
Chen W~K, Liu X~Y, Fang W~H, Dral P~O and Cui G 2018 {\em J. Phys. Chem.
  Lett.\/} {\bf 9} 6702--6708

\bibitem{Guan2019PCCP}
Guan Y, Zhang D~H, Guo H and Yarkony D~R 2019 {\em Phys. Chem. Chem. Phys.\/}
  {\bf 21}(26) 14205--14213

\bibitem{Westermayr2019CS}
Westermayr J, Gastegger M, Menger M~F~S~J, Mai S, González L and Marquetand P
  2019 {\em Chem. Sci.\/} {\bf 10}(35) 8100--8107

\bibitem{Nanbu2010CS}
Nanbu S, Ishida T and Nakamura H 2010 {\em Chem. Sci.\/} {\bf 1}(6) 663--674

\bibitem{Harris1989}
Harris D~C and Bertolucci M~D 1989 {\em Symmetry and spectroscopy: an
  introduction to vibrational and electronic spectroscopy\/} (New York: Dover
  Publications)

\bibitem{Ramakrishnan2014SD}
Ramakrishnan R, Dral P~O, Rupp M and von Lilienfeld O~A 2014 {\em Sci. Data\/}
  {\bf 1}

\bibitem{Huang2016JCP}
Huang B and von Lilienfeld O~A 2016 {\em J. Chem. Phys.\/} {\bf 145} 161102

\bibitem{Schuett2018JCP}
Sch\"{u}tt K~T, Sauceda H~E, Kindermans P~J, Tkatchenko A and M\"{u}ller K~R
  2018 {\em J. Chem. Phys.\/} {\bf 148} 241722

\bibitem{Nebgen2018JCTC}
Nebgen B, Lubbers N, Smith J~S, Sifain A~E, Lokhov A, Isayev O, Roitberg A~E,
  Barros K and Tretiak S 2018 {\em Journal of Chemical Theory and
  Computation\/} {\bf 14} 4687--4698

\bibitem{Sifain2018JPCL}
Sifain A~E, Lubbers N, Nebgen B~T, Smith J~S, Lokhov A~Y, Isayev O, Roitberg
  A~E, Barros K and Tretiak S 2018 {\em J. Phys. Chem. Lett.\/} {\bf 9}
  4495--4501

\bibitem{pereira2018JC}
Pereira F and Aires-de Sousa J 2018 {\em J. Cheminf.\/} {\bf 10} 43 ISSN
  1758-2946

\bibitem{Schuett2019}
Sch{\"u}tt K~T, Gastegger M, Tkatchenko A and M{\"u}ller K~R 2019 {\em
  Quantum-Chemical Insights from Interpretable Atomistic Neural Networks\/}
  (Springer International Publishing) pp 311--330

\bibitem{Schuett2019JCTC}
Sch\"utt K~T, Kessel P, Gastegger M, Nicoli K~A, Tkatchenko A and M\"uller K~R
  2019 {\em J. Chem. Theory Comput.\/} {\bf 15} 448--455

\bibitem{Christensen2019JCP}
Christensen A~S, Faber F~A and von Lilienfeld O~A 2019 {\em J. Chem. Phys.\/}
  {\bf 150} 064105

\bibitem{Corinna1994}
Cortes C, Jackel L~D, Solla S~A, Vapnik V and Denker J~S 1994 Learning curves:
  Asymptotic values and rate of convergence {\em Advances in Neural Information
  Processing Systems 6\/} ed Cowan J~D, Tesauro G and Alspector J
  (Morgan-Kaufmann) pp 327--334

\bibitem{Mueller1996NC}
M\"{u}ller K~R, Finke M, Murata N, Schulten K and Amari S 1996 {\em Neural
  Comput.\/} {\bf 8} 1085--1106

\bibitem{vonLilienfeld2018ACIE}
{von Lilienfeld} O~A 2018 {\em Angew. Chem. Int. Edit.\/} {\bf 57} 4164--4169

\bibitem{Schuett2019NC}
Sch\"{u}tt K~T, Gastegger M, Tkatchenko A, M\"{u}ller K~R and Maurer R~J 2019
  {\em Nat. Commun.\/} {\bf 10} 5024

\bibitem{Plasser2016JCTC}
Plasser F, Ruckenbauer M, Mai S, Oppel M, Marquetand P and Gonz\'alez L 2016
  {\em J. Chem. Theory Comput.\/} {\bf 12} 1207

\bibitem{Akimov2018JPCL}
Akimov A~V 2018 {\em J. Phys. Chem. Lett.\/} {\bf 9} 6096--6102

\bibitem{sharc-md2}
Mai S, Richter M, Ruckenbauer M, Oppel M, Marquetand P and Gonz\'alez L 2018
  Sharc2.0: Surface hopping including arbitrary couplings -- program package
  for non-adiabatic dynamics sharc-md.org

\bibitem{Wigner1932PR}
Wigner E 1932 {\em Phys. Rev.\/} {\bf 40}(5) 749--750

\bibitem{QML}
Christensen A, Faber F, Huang B, Bratholm L, Tkatchenko A, M\"{u}ller K and
  Lilienfeld O 2017 Qml: A python toolkit for quantum machine learning
  https://github.com/qmlcode/qml

\bibitem{Gastegger2015JCTC}
Gastegger M and Marquetand P 2015 {\em J. Chem. Theory Comput.\/} {\bf 11}
  2187--2198

\bibitem{Chmiela2018NC}
Chmiela S, Sauceda H~E, M\"{u}ller K~R and Tkatchenko A 2018 {\em Nat.
  Commun.\/} {\bf 9} 3887

\bibitem{Faber2018JCP}
Faber F~A, Christensen A~S, Huang B and von Lilienfeld O~A 2018 {\em J. Chem.
  Phys.\/} {\bf 148} 241717

\bibitem{Christensen2019arXiv}
Christensen A~S, Bratholm L~A, Faber F~A, Glowacki D~R and von Lilienfeld O~A
  2019 Fchl revisited: faster and more accurate quantum machine learning

\bibitem{Faber2017JCTC}
Faber F~A, Hutchison L, Huang B, Gilmer J, Schoenholz S~S, Dahl G~E, Vinyals O,
  Kearnes S, Riley P~F and von Lilienfeld O~A 2017 {\em J. Chem. Theory
  Comput.\/} {\bf 13} 5255--5264

\bibitem{Walt2011CSE}
van~der Walt S, Colbert S~C and Varoquaux G 2011 {\em Comput. Sci. Eng.\/} {\bf
  13} 22--30 ISSN 1521-9615

\bibitem{TDT2016a}
{Theano Development Team} 2016 {\em arXiv\/} {\bf abs/1605.02688}

\bibitem{Adam2014}
Kingma D~P and Ba J 2014 {\em arXiv:1412.6980\/} {\bf abs/1412.6980}

\bibitem{Ruder2017arXiv}
Ruder S 2017 {\em arXiv:1706.05098\/} {\bf abs/1706.05098}

\bibitem{Artrith2011PRB}
Artrith N, Morawietz T and Behler J 2011 {\em Phys. Rev. B\/} {\bf 83} 153101

\bibitem{Yao2018Cs}
Yao K, Herr J~E, Toth D, Mckintyre R and Parkhill J 2018 {\em Chem. Sci.\/}
  {\bf 9}(8) 2261--2269

\bibitem{Marquetand2004JCP}
Marquetand P, Materny A, Henriksen N~E and Engel V 2004 {\em J. Chem. Phys.\/}
  {\bf 120} 5871--5874

\bibitem{Bonafe2018JPCL}
Bonaf\'{e} F~P, Hern\'{a}ndez F~J, Aradi B, Frauenheim T and S\'{a}nchez C~G
  2018 {\em J. Phys. Chem. Lett.\/} {\bf 9} 4355--4359

\bibitem{Cheng2016arXiv}
Cheng H~T, Koc L, Harmsen J, Shaked T, Chandra T, Aradhye H, Anderson G,
  Corrado G, Chai W, Ispir M, Anil R, Haque Z, Hong L, Jain V, Liu X and Shah H
  2016 {\em arXiv:1606.07792\/}

\end{thebibliography}
\providecommand{\newblock}{}

%

\clearpage 
\title{Supporting Information}
\section{Machine Learning (ML) models}

\subsection{Kernel Ridge Regression (KRR)}

For Kernel Ridge Regression (KRR) we used the QML toolkit~\cite{QML} and as a representation, FCHL18, the inverse distance matrix (inv.D.), and a development version of FCHL19 were tested. For FCHL a gaussian kernel and default settings were taken if not mentioned otherwise. The cut-off distance was set to 20 \AA ~in order to capture the whole molecule. A hyperparameter search was conducted with focus on two main parameters: the kernel width, $\sigma$, and $\lambda$, a hyperparameter that is responsible for the strength of regularization. For all models, the regularization $\lambda$ was set to $10^{-9}$. For the inv.D. matrix, a Gaussian kernel with $\sigma=0.65$ turned out to give best results for the single-state models. When we learn the nonadiabatic couplings multiplied with the corresponding energy gap, we use a $\sigma$ of 1.5, the same value was used for FCHL for this model. The laplacian kernel instead of Gaussian kernel in combination with the inverse distance matrix resulted in a negative correlation. When including the state representation, $\sigma$ was set to 10 and $\sigma_{state}$ to 0.2. For the FCHL representation, $\sigma$ was set to 4.0, 0.8, and 0.65 for the single-state model, the model that treats gradients as derivatives and for the state model in combination with $\sigma_{state}=1.2$, respectively. A Laplacian kernel was tested too, also for encoding the quantum energy level, but did not improve results.

\section{Multi-layer feed-forward neural networks (NNs)}
Parameters for the NN-models implemented using theano~\cite{TDT2016a} are given in Table \ref{tab:param}. The number of hidden layers was 6 and 50 nodes per hidden layer were used for all NN/inv.D. and NN-All models. The number of hidden layers was converged at 4 for the NN/FCHL model that treats each property separately. As basis functions we used the hyperbolic tangent for all models except for those trained on energies (and gradients), where we used the shifted softplus function, $ln(0.5 e^{x}+0.5)$. 
Table \ref{tab:param} shows the chosen hyperparameters for NNs. The L2 regularization was set to $1\cdot 10^{-9}$ if not stated otherwise and the batch size was set to 10 or 50, depending on the number of data points used for training.

\subsection{NN-All/inv.D.}
\begin{figure}[h]
    \centering
    \includegraphics[scale=0.5]{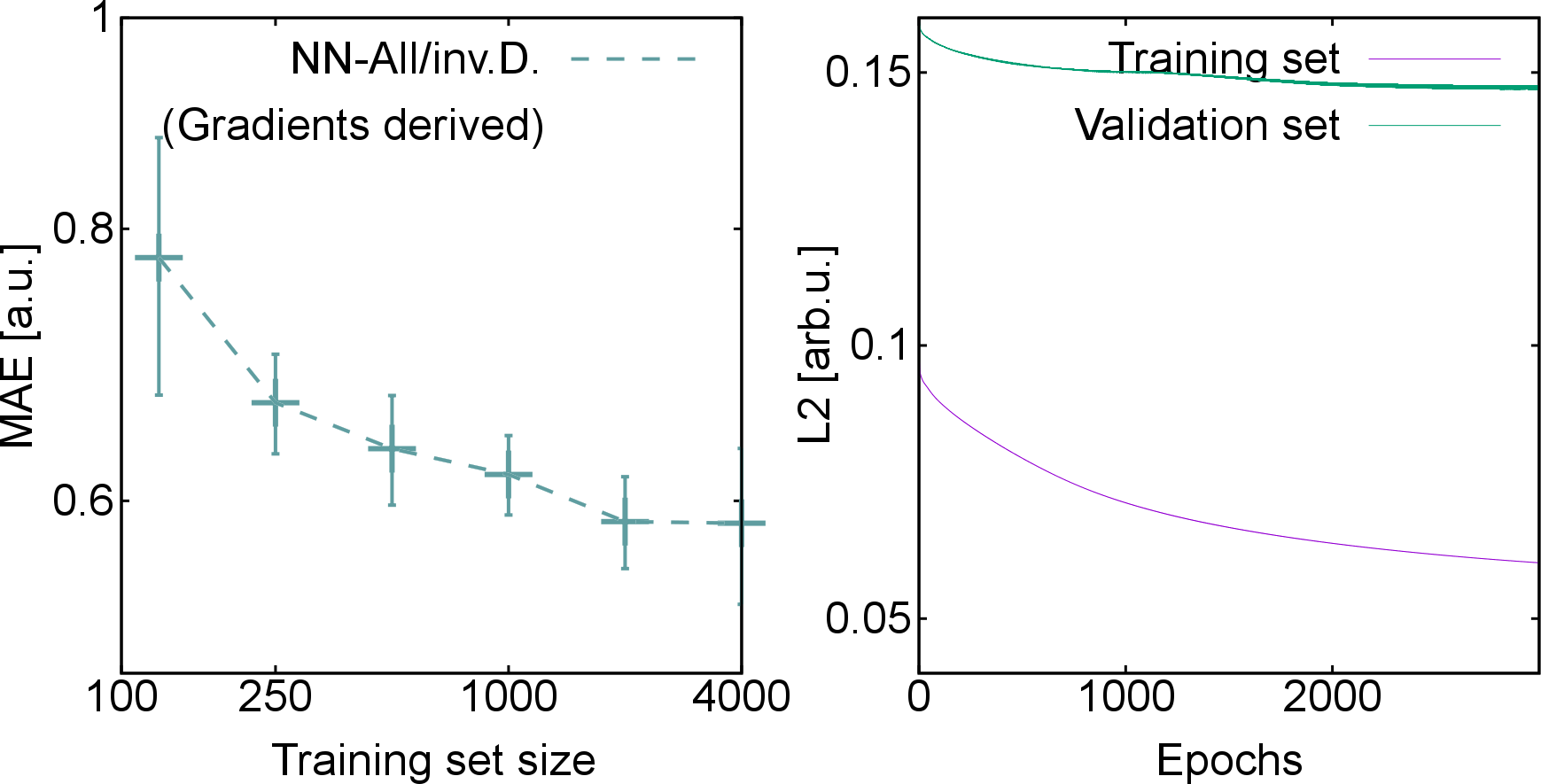}
    \caption{Results of the NN-All/inv.D. model in addition to Fig. 2 in the main text. The learning curves are obtained by summing up all MAEs of all properties in a.u. and are shown in the left plot with the corresponding loss functions on the training and validation set for a model trained on 4000 data points (3600 training, 400 validation) in the right plot.}
    \label{NN-All}
\end{figure}

The NN-All/inv.D. model that treats gradient as derivatives of ML potentials for energies shows "anti"-learning curves when solely taking a look at the gradients. The overall learning curve summed from all properties is shown in Fig.~\ref{NN-All} and clearly shows the overall negative slope. Also the loss function indicates successful learning of the model.
{}

\begin{table*}[h]
\caption{\label{tab:param}Parameters for different NN models with the inverse distance matrix (inv.D.) and the FCHL representation. "E" is used for models trained on energies, "E+G" abbreviates models trained on energies and gradients together, "E$^{states}$" indicate models that explicitly include the energetic state, "G" indicates gradients, "$\mu$" permanent and transition dipole moments and "NACs" gives parameters for models trained on nonadiabatic couplings, once without (NACs) and once with (NACs ($\Delta E$)) including physical relations.}\begin{tabular}{llllllll}
\hline
\textbf{NN/inv.D.} &E &E$^{states}$  &E+G  &G &NACs &NACs ($\Delta E$)&$\mu$\\
\hline 
Learning rate, $lr$ &$4.3\cdot 10^{-2}$     &$2.1\cdot 10^{-4}$     &$1.0\cdot 10^{-3}$     &4.3$\cdot 10^{-2}$     &$6.1\cdot 10^{-4}$ &8.4$\cdot 10^{-3}$ &$9.0\cdot 10^{-4}$\\
Factor to anneal $lr$ &- &0.965  &999 &- &0.963&0.978 &0.992 \\
Steps for annealing $lr$ &- &15 &64 &- &11 &49 &58 \\
L2 regularization & & &1.2$\cdot 10^{-10}$ & &1.3$\cdot 10^{-7}$ &1.3$\cdot 10^{-7}$&5.0$\cdot 10^{-8}$\\
\hline 
\hline
\textbf{NN/FCHL} &E &E$^{states}$   &G &NACs &NACs ($\Delta E$)&$\mu$\\
\hline 
Learning rate, $lr$ &2.5$\cdot 10^{-4}$     &7.8$\cdot 10^{-5}$     &1.6$\cdot 10^{-4}$     &6.5$\cdot 10^{-3}$     &$6.1\cdot 10^{-3}$ &$1.5\cdot 10^{-3}$\\
Factor to anneal $lr$ &0.985 &0.959  &0.983 &0.970 &0.980 &0.994 \\
Steps for annealing $lr$ &87 &70&62 &54 &70 &3 \\
L2 regularization & &7.1$\cdot 10^{-9}$ \\

\hline
\hline
\textbf{NN-All} &inv.D. &inv.D. (+G) &FCHL\\
\hline
Learning rate, $lr$ &$4.3\cdot 10^{-3}$&7.2$\cdot 10^{-5}$ &6.3$\cdot 10^{-6}$	\\
Factor to anneal $lr$ &0.970 &0.997&0.985 \\
Steps for annealing $lr$ 	&54 &59 &87\\
\end{tabular}
\centering
\end{table*}
\end{document}